\documentclass[a4paper,11pt]{article}
\usepackage{graphicx} 
\usepackage{bbold}
\usepackage{mathtools}
\usepackage{amsmath}
\usepackage{amsfonts}
\usepackage{amssymb}
\usepackage{slashed} 
\usepackage{color}
\usepackage[table]{xcolor}
\usepackage{hyperref}
\usepackage{bm}
\usepackage{mathrsfs}
\usepackage{cite}
\usepackage{physics}
\def\eq#1{{Eq.~(\ref{#1})}}

\def\abs#1{\left| #1\right|}

\colorlet{grayline}{gray!70}
\definecolor{mf}{rgb}{204,0,0}
\definecolor{blueline}{rgb}{0,0.27,0.55}
\definecolor{DarkGray}{gray}{0.4}
\definecolor{Gray}{gray}{0.6}
\definecolor{oucrimsonred}{rgb}{0.6, 0.0, 0.0}
\definecolor{persianblue}{rgb}{0.11, 0.22, 0.73}
\definecolor{forestgreen}{rgb}{0.13,0.35,0.13}
 \hypersetup{colorlinks, citecolor=forestgreen, linkcolor=forestgreen, urlcolor=forestgreen}
\newcommand{\be}{\begin{equation}}
\newcommand{\ee}{\end{equation}}
\newcommand{\bea}{\begin{eqnarray}}
\newcommand{\eea}{\end{eqnarray}}

\newcommand*\xbar[1]{%
  \hbox{\;%
    \vbox{%
      \hrule height 0.5pt 
      \kern0.5ex
      \hbox{%
        \kern-0.25em
        \ensuremath{#1}%
        \kern-0.07em
      }%
    }%
  }%
} 
\newcommand{\com}[1]{}
\newcommand{\gsim}{\lower.7ex\hbox{$\;\stackrel{\textstyle>}{\sim}\;$}}
\newcommand{\lsim}{\lower.7ex\hbox{$\;\stackrel{\textstyle<}{\sim}\;$}} 

\newcommand{\bc}{\begin{center}}
\newcommand{\ec}{\end{center}}

\newcommand{\K}{K^{*}(892)^0}

\newcommand{\lambdaA}{\lambda_{1}}
\newcommand{\lambdaB}{\lambda_{2}}
\newcommand{\lambdaAp}{\lambda_1^{\prime}}
\newcommand{\lambdaBp}{\lambda_2^{\prime}}

\newcommand{\JP}{J/\psi}

\newcommand{\Bell}{{\cal B}}
\newcommand{\II}{{\cal I}_3}
\font\beeg=cmr17 scaled 1800
\newbox\ibox
\def\versal#1{\setbox\ibox=\hbox{{\beeg #1}~}
	    \noindent\global\hangindent=\wd\ibox\global\hangafter-2
	    \sc\smash{\llap {\lower 14pt \box\ibox}}}

\begin{document}

\thispagestyle{empty}
\begin{center}
{\color{oucrimsonred}\Large {\bf  Entanglement and Bell inequality violation in $B\to\phi\phi$ decays}}

\vspace*{1.5cm}
{\color{black}
{\bf E. Gabrielli$^{{a,b,c}}$} and {\bf L. Marzola$^{{c,d}}$}
}\\

\vspace{0.5cm}
       {\small
           {\it \color{black}
    (a) Physics Department, University of Trieste, Strada Costiera 11, \\ I-34151 Trieste, Italy}
  \\[1mm]  
{\it  \color{black}
(b) INFN, Sezione di Trieste, Via Valerio 2, I-34127 Trieste, Italy}
    \\[1mm]
  {\it \color{black}
(c) Laboratory of High-Energy and Computational Physics, NICPB, R\"avala pst 10, \\ 10143 Tallinn, Estonia}
}
\\[1mm]
{\it \color{black}
(d) Institute of Computer Science, University of Tartu, 
Narva mnt 18, \\ 51009 Tartu, Estonia.}

\ec

 \vskip0.5cm
\bc
{\color{black}
\rule{0.7\textwidth}{0.5pt}}
\ec
\vskip1cm
\bc
{\bf ABSTRACT}
\ec

   \noindent
The decays of the $B$ meson into vector mesons, observed at the LHCb experiment, provide an ideal laboratory to investigate  particle physics phenomena with quantum information theory methods. In this article, we focus on the decays yielding a pair of $\phi$  mesons to investigate the presence of entanglement in the spin correlations of the system and quantify the amount of Bell inequality violation it entails. Our results show that the present LHCb data allows access to entanglement and to the Bell inequality violation with a significance exceeding the 5$\sigma$ threshold in both the cases.  This demonstrates that the strong and electroweak interactions responsible for the $B$ meson decay act as a source of entanglement and the quantum mechanics nature of high-energy phenomena. Particular attention is paid to the assessment of loopholes: deficiencies in the experimental setup which could invalidate the results of the Bell test.

\vspace*{5mm}

\newpage
\pagestyle{plain}

\section{ Introduction}

One of the most fascinating aspects of quantum mechanics (QM) is the possible presence of entanglement in composite systems~\cite{Amico20081,Horodecki:2009zz,Guhne20091,Laflorencie20161}, a phenomenon that has no counterpart in classical or fully deterministic theories. The latter, in particular, are built on the assumption of {\it local realism}, which endows physical systems with intrinsic properties that exist whether or not we perform a measurement (realism). The concept of {\it locality} is here understood in the sense that a physical system can only be influenced by its local surroundings and is causally connected to other systems only through interactions that do not propagate faster than the speed of light. 

The amazing property of entangled systems is that entanglement keeps linking the composing subsystems even if these are separated at an arbitrary distance. This connection maintains the overall quantum system as a whole, demonstrating the true {\it non-local} nature of quantum phenomena. For instance, if we consider a system formed by two entangled particles and perform a measurement on one of these, the other one -- no matter how far away -- will collapse instantaneously into a new state selected by the outcome of the measurement performed on the first particle. In other words, once quantum systems are entangled, the state describing the composite system cannot be reconstructed from the states describing the subsystems in isolation. Mathematically, the idea that entangled systems are more than the sum of their parts is reflected in the fact that the corresponding state cannot be written as the tensor product of the quantum states describing the separate subsystems. Clearly, the contrary holds within theories based on local realism as all the properties of the system and of its subsystems are predetermined and exist prior to any measurement.

In 1935, Einstein, Podolsky and Rosen suggested a {\it gedanken} experiment~\cite{Einstein:1935rr} which showed that the description of reality based on the wave function of QM is incomplete if local realism is assumed. This suggested the idea that the incompleteness of QM could be resolved in the framework of {\it local hidden variables} (LHV) theories, which  explain the apparent probabilistic nature of quantum phenomena by introducing additional (perhaps experimentally unaccessible) variables -- we refer the Reader to Ref.~\cite{Genovese:2005nw} for a recent review on the topic.

The diatribe between QM and LHV theories entered a new phase in 1969, when John Bell devised a test able to actually discriminate between the two frameworks. The idea is based on correlated measurements performed independently on the spins of two spatially separated parts of an entangled system~\cite{Bell:1964,bell2004speakable}. Assuming that these measurements do not affect each other and are mutually independent yields an upper bound on the set of expectation values of correlated measurements, encoded in the so-called \textit{Bell inequality}~\cite{Bell:1964,bell2004speakable}.\footnote{Different forms of the inequality tailored to systems with different dimensionalities have been proposed in Refs.~\cite{Horodecki:1995340,Brunner:RevModPhys.86.419,Collins:2002sun,Kaszlikowski:PhysRevA.65.032118}.} In the test, the intrinsic {\it non-local} nature of QM shows in correlations which are stronger than the ones expected within any deterministic or LHV theory. As a consequence, QM can violate the bound posed by the Bell inequality and experimental measurements of these correlation can then rule out competing interpretations based on local realism~\cite{bell2004speakable,bertlmann2016quantum}.

The possible violation of the Bell inequality was investigated in experiments with entangled photons, running at energies of the order of few eV~\cite{Horodecki:2009zz}. In these tests, a system composed by two photons is prepared into a singlet spin state and the spin of each photon is subsequently measured along different directions. This allows the experimental reconstruction of the correlations used in the Bell test, giving access to entanglement and to the violation of the Bell inequality that it may entail. The violation has been verified for the first time in optical experiments~\cite{Aspect:1982fx,Weihs:1998gy} and a large effort was devoted to close -- almost all -- potential loopholes in low-energy tests with photons~\cite{Hensen:2015ccp,Giustina:2015yza} and with atoms~\cite{Rosenfeld:2017rka}.

Simply put, loopholes are deficiencies in the experimental apparatus that can be exploited to bypass the consequences of Bell's theorem, thereby invalidating the conclusions of the corresponding test. The most common loophole affecting the setups relevant to our study is the {\it locality} one, which exploits the possible time-like separation among the two observers involved in the test to claim an exchange of information and the consequent loss of independence in the performed measurements. Indeed, the effect of a genuine non-locality can be mimicked by local interactions if the choice of setting at a measurement site is influenced by the result obtained by the other observer. Other loopholes relevant in this context are those of {\it detection}~\cite{Pearle:1970zt}, {\it coincidence}~\cite{Larsson:2003efo}, {\it freedom of choice}~\cite{Bohm:1957zz} and {\it super determinism}~\cite{Larsson_2014}. We refer the reader to Ref.~\cite{Barr:2024djo} for more details.   
 
In the context of particle physics, entanglement has been probed using low-energy protons~\cite{Lamehi-Rachti:1976wey}, while possible tests at high energy colliders were first proposed in~\cite{Tornqvist:1980af,Privitera:1991nz,Abel:1992kz}. Other entanglement and Bell inequality tests have been also suggested in the framework of positronium~\cite{Acin:2000cs,Li:2008dk}, and charmonium decays~\cite{Baranov:2008zzb,Chen:2013epa,Qian:2020ini}, as well as with neutrino oscillations~\cite{Banerjee:2015mha} and with neutral meson~\cite{Benatti2000,Bertlmann:2001ea,Banerjee:2014vga,Go:2003tx}, although in this last case only an indirect test of Bell inequality violation could be probed.

The interest in the high energy phenomenology of entanglement 
was recently revived after it was shown that entanglement could be observed with the spin correlations of top-quark pairs produced at the LHC~\cite{Afik:2020onf} and that the Bell inequality violation can also be accessible through the same system~\cite{Fabbrichesi:2021npl}. More works on these topic followed, studying in particular the top-quark production~\cite{Severi:2021cnj,Aguilar-Saavedra:2022uye,Larkoski:2022lmv,Afik:2022kwm,Han:2023fci}, hyperons~\cite{Gong:2021bcp,Fabbrichesi:2024rec}, tau-pairs~\cite{Ehataht:2023zzt,Fabbrichesi:2024wcd}, and gauge bosons production from Higgs boson decay~\cite{Barr:2021zcp,Ashby-Pickering:2022umy,Aguilar-Saavedra:2022wam,Fabbrichesi:2023cev} and in vector boson scattering~\cite{Morales:2023gow}. These studies collectively show that entanglement and the Bell inequality violation are accessible in several systems produced at high energy collider experiments and that the related phenomenology can be used to probe physics beyond the Standard Model~\cite{Fabbrichesi:2022ovb,Maltoni:2024tul,Fabbrichesi:2023jep,Aoude:2023hxv,Bernal:2023ruk, Fabbrichesi:2024xtq, Fabbrichesi:2024wcd }. 

The possibility of detecting entanglement was recently verified by the ATLAS~\cite{ATLAS:2023fsd} and CMS~\cite{CMS:2024pts} collaborations at CERN, which observed its presence with a significance of more than $5\sigma$ in the spin correlations of top-quark pairs produced near threshold at the LHC. The violation of the Bell inequality at high energy was instead verified for the first time by analyzing the data pertaining to the $B$ meson decays into two spin-1 mesons~\cite{Fabbrichesi:2023idl} gathered by the LHCb collaboration~\cite{LHCb:2013vga}. The result firmly establishes the presence of this quantum mechanical hallmark with a significance exceeding the $5\sigma$ level for a bipartite system formed by two qutrits (three-level systems), at energies of the order of 5 GeV--a billion times larger than those utilized in optical experiment~\cite{Aspect:1982fx,Weihs:1998gy,Hensen:2015ccp} -- and in the presence of strong and weak interactions. In regard of this, the $B^0_d\to \JP~\K$ decay used in the study is vulnerable to the {\it locality} loophole, brought about in this case by the difference between the lifetimes of the $\JP$ and $\K$ mesons. The loophole uses the fact that the measurements of the spin of each subsystem -- identified here with the decay of the particles, which reveals the orientation of the decaying particle spin vector -- are time-like separated. Since potential information transfer among the two subsystems is then possible, the hypothesis of independent measurements underlying the Bell test may be argued to be consequently not respected. 

In this article we then review the Bell inequality violation in the $B$ meson decays focussing on a similar process, the $B^0_s\to \phi \phi$ decay, where the locality loophole can be kept under control. As we will see, although the significance of the violation is lower than for the $B^0_d\to \JP~\K$ decay, the presence of identical particles in the final state makes communication between the two parties impossible in a large majority of the recorded events. In fact, by performing $10^5$ pseudo-experiments that mimic the decay process of interest, we find that almost 95\% of the $\phi$ pair decays take place at space-like separations. This ensures that the large majority of the samples analyzed in actual experiments can be safely used to perform a Bell test as every form of local communication between the decaying particles is automatically prevented.

\section{Quantum observables}
We start by summarizing the main results pertaining to the measurement of spin correlations in the bipartite qutrit systems created in the decays of a spin-0 particle. For the Bell test, we use a specific inequality tailored to this system. To quantify the amount of entanglement in the spin correlation, instead, we rely on the entropy of entanglement~\cite{Horodecki:2009zz} as the vanishing spin of the decaying particle forces the resulting bipartite system to be in a pure state~\cite{Fabbrichesi:2024rec}. Central to the analysis is the helicity density matrix, $\rho$, which we introduce in the forthcoming section.

\subsection{Density operator, polarization and spin correlations}

A quantum system whose state $\ket{\psi}$ is known exactly is said to be in a  {\it pure state}. The description of {\it mixed systems}, on the contrary, uses an ensemble $\{(p_i,\ket{\psi_i})\}$ of pure states $\ket{\psi_i}$ and  associated probabilities $p_i$ that encode our incomplete knowledge of the system. From this, we built an operator 
\begin{equation}
  \rho = \sum_i p_i \op{\psi_i}\,,\qquad \sum_i p_i = 1\,,
\end{equation} 
which qualifies as a \textit{density operator} if and only if
\begin{enumerate}
  \item[i)] $\tr(\rho)=1$
  \item[ii)] $\rho\geq0$, which implies $\rho=\rho^\dagger$.
\end{enumerate} 

The density operator of a pure state is simply given by $\rho=\op{\psi}$, as we know exactly which of the states $\{\ket{\psi_i}\}$ actually describes the system. It then follows that for pure states
\begin{equation}
  \rho^2 = \rho\,,
\end{equation}
as the density operator coincides with the $\ket{\psi}$ subspace projector.  Notice that in general $\tr(\rho^2)\leq1$, with the equality holding only for pure states. 

In the following, we aim to describe the polarization state of two spin-1 particles of arbitrary non vanishing masses. Admitting three degenerate spin -- or, equivalently, helicity -- states, these particles are prototypal examples of the qutrits of quantum information theory. A state $\ket{\psi_i}$, describing one of these qutrits will then be a coherent superposition of the three basis vectors representing the (orthogonal) possible polarizations $\lambda = +1,\, 0,\,-1$ of the system. The density operator describing the polarization state of a massive spin one particle is that of a qutrit and, as such, can be represented with a $3\times3$ matrix depending on 8 parameters. This can be decomposed as
\begin{equation}
  \rho_{\lambdaA, \lambdaAp} = \frac{1}{3} \mathbb{1}_3 + \sum_{a=1}^8 v_a T^a\,,
  \quad\text{ with }\,
  \lambdaA,\lambdaAp = +1,\, 0,\,-1\,,
\end{equation} 
on the basis formed by the Gell-Mann matrices $T^a$, with $a=1,\dots,8$, and by the  $3\times3$ unit matrix $\mathbb{1}_3$. The coefficients $v_a$ are necessarily real. Similarly, once explicitly written in matrix representation, the density operator describing the joint state of two qutrits is a $9\times9$ matrix with indices $\lambdaA\otimes\lambdaB$, $\lambdaAp\otimes\lambdaBp$\footnote{We indicate with a tensor product of helicity labels the label of the basis vector given by the tensor product of the corresponding helicity states. These products map to Kronecker products of the basis vectors used in the chosen  representation of the helicity states as $|\psi\rangle \otimes |\phi\rangle = (\psi_1 |\phi\rangle,\psi_2 |\phi\rangle,\dots\psi_n |\phi\rangle)^T = (\psi_1 \phi_1 ,\psi_1 \phi_2 \dots \psi_n \phi_n)^T$ for two helicity states $|\psi\rangle$ and $|\phi\rangle$ with components $\psi_i$ and $\phi_i$, respectively.} . As such, it can be written as a sum of tensor products involving the Gell-Mann matrices, $T^a$ with $a=1,\dots,8$ and the $3\times3$ unit matrix $\mathbb{1}_3$ as follows
\bea
\label{eq:rhone}
\rho_{\lambdaA\otimes\lambdaB, \lambdaAp\otimes\lambdaBp}
= \Big[
  \frac{\mathbb{1}_3\otimes
  \mathbb{1}_3}{9}
    +
    \sum_{a=1}^8 f_a (T^a \otimes \mathbb{1}_3)
    +
    \sum_{a=1}^8 g_a (\mathbb{1}_3\otimes T^a) 
    +\sum_{a,b=1}^8 h_{ab}  (T^a\otimes T^b)\Big]_{\lambdaA\otimes\lambdaB, \lambdaAp\otimes\lambdaBp}\, .
\label{rho}
\eea
The coefficients $f_a$ and $g_a$, $a=1,\dots,8$, as well as the 64 elements of the symmetric matrix $h_{ab}$, generally depend on the kinematics of the process yielding the production of the two qutrits and can be computed analytically from the related transition amplitudes. The corresponding averages can be experimentally reconstructed typically from the angular distributions of the decay products of the same qutrits, which indicate the orientation of the spin vectors of the latter. Complete quantum tomography is achieved once these averages are known and the full density matrix describing the sample of qutrit pairs under scrutiny is thus determined.

These coefficients can be obtained by projecting $\rho$ on the desired subspace basis via the traces
$$f_a=\frac{1}{6}\,\tr\left[\rho \left(T^a \otimes \mathbb{1}_3\right)\right]\, , ~~
g_a=\frac{1}{6}\,\tr\left[\rho \left(\mathbb{1}_3\otimes T^a\right)\right]\, , ~~
h_{ab}=\frac{1}{4}\,\tr\left[\rho \left(T^a \otimes T^b\right)\right]\,.$$
The averages $\ev{f^a}$ and $\ev{g^a}$ determine the \textit{linear and tensor polarizations}\footnote{These quantities coincide with the linear and tensor polarizations of the two qutrit sample if the density matrix in \eq{eq:rhone} is expanded on the basis formed by the tensor product of irreducible tensor operators formed with the spin-1 $3\times3$ matrices.} of the qutrit pair sample created in the experiment; the matrix $\ev{h^{ab}}$ contains instead the \textit{spin correlations} that may contain entanglement and lead to the violation of the Bell inequality.

If we now define ${\cal M}(\lambdaA,\lambdaB)$ the matrix element for the transition amplitude yielding two spin-one particles with helicities $\lambdaA$ and $\lambdaB$, the helicity density operator describing the spin state of the resulting bipartite system can be computed as
\bea
\rho(\lambdaA,\lambdaAp,\lambdaB,\lambdaBp)&=&\frac{
{\cal M}(\lambdaA,\lambdaB) {\cal M}^\dag (\lambdaAp,\lambdaBp)}{{|\cal M}|^2}\, ,
\label{rho2}
\eea
where, as usual, $|{\cal M}|^2$ stands for the unpolarized square amplitude obtained by summing over the polarizations. A sum over possible internal degrees of freedom of initial state particles, including the spin, is also understood in both the numerator and the denominator.

\subsection{Connection with the helicity amplitude spin formalism}

The \textit{helicity amplitudes} are the matrix elements of the $S$ matrix taken between initial and final helicity states. The latter are usually defined in the center-of-mass frame where the scattering angle also indicates the quantization axis used to define the helicity states $\ket{\lambda_i}$ of the final state particles, which recover the usual spin states once boosted to a frame where the $i$-th particle is at rest. To understand how the spin formalism of helicity amplitudes is particularly suitable for the computation of the helicity density matrix, consider the matrix element ${\cal M}(\lambdaA,\lambdaB,\chi_1, \chi_2)$ for a $2\to2$ reaction. The symbols $\chi_1$ and $\chi_2$ indicate the helicities of the initial state particles and, as before, $\lambdaA$ and $\lambdaB$ those of the final state ones. By decomposing the $S$ matrix as $S=1+iT$, we can write 
\begin{equation}
  \delta^{(4)}(p_1+p_2-k_1-k_2){\cal M}(\lambdaA,\lambdaB) \propto \mel{\Omega(\theta,\phi), \lambdaA\, \lambdaB}{T}{\Omega(0,0), \chi_1\,\chi_2}\,,
\end{equation}
where $k_1$ and $k_2$ are the initial state particle momenta that we take along the $z=\Omega(\theta=0,\phi=0)$ direction, with $\theta$ and $\phi$ being the polar and azimuthal angles. The transition amplitude can then be expanded into partial-wave amplitudes to give
\begin{equation}
  \mel{\Omega(\theta,\phi), \lambdaA\, \lambdaB}{T}{\Omega(0,0), \chi_1\,\chi_2}
  = \frac{1}{4\pi} \sum_J (2J+1)\mel{\lambdaA \lambdaB}{T^J}{\chi_1\chi_2}\mathcal{D}^{J*}_{\chi \lambda}(\phi, \theta,0)\,,
\end{equation}   
where $\mathcal{D}^{J}_{\chi, \lambda}$ are the Wigner $D$-matrix elements of the spin-$J$ representation of the rotation group, $\chi=\chi_1-\chi_2$ and $\lambda = \lambdaA-\lambdaB$. The helicity density matrix can then be determined as 
\begin{align}
  \label{eq:rhoHA}
\rho_{\lambdaA\otimes\lambdaB, \lambdaAp\otimes\lambdaBp}
& =
\frac{1}{\abs{\mathcal{M}}^2} 
\sum_J w_{\lambdaA\lambdaB}^J\,w^{*J}_{\lambdaAp\lambdaBp} \sum_{k=-J}^J  \mathcal{D}^{J*}_{k, \lambdaA-\lambdaB}(0, \theta,0) \mathcal{D}^{J}_{k, \lambdaAp-\lambdaBp}(0, \theta,0)
\end{align}
where a sum over the helicities of the initial state is understood and $w_{\lambdaA\lambdaB}^J\propto \mel{\lambdaA \lambdaB}{T^J}{\chi_1\chi_2}$.  The overall factor is set by the condition $\tr(\rho)=1$ and the dependence on the azimuthal angle $\phi$ drops out as required by the cylindrical symmetry enjoyed by the process. A similar result holds for  $1\to2$ processes, where $J$ is identified with the spin of the decaying particle.   

For the considered case of a (pseudo)scalar decaying into two massive spin-1 particles, as in the $B\to\phi\phi$ process, the calculation of \eq{eq:rhoHA} simplifies considerably.  The fact that the $B^0_s$ meson has spin 0 removes any dependence on the scattering angle $\theta$ from the density matrix. In fact, the non-vanishing $D$-matrix element are forced to unit and we are left only with three non-vanishing helicity amplitudes (as also required by angular momentum  conservation):
\be
w_{\lambda\lambda^\prime} \propto \mel{\lambda \lambda'}{\mathcal{H}}{B}\propto \langle V_{1}(s_1=\lambda) V_{2}(s_2=-\lambda^\prime)| {\cal H} |B\rangle\,\quad \text{with} \quad \lambda=\lambda^\prime=+1,\, 0,\text{ or }\,-1\,.
\ee
In the above formula we indicated with ${\cal H}$ the interaction Hamiltonian responsible for the transition and with $s_i$ the spin of the particle $V_i$, as measured in the corresponding rest frame and taking as quantization axis the direction of motion of one of the two particles in the $B^0_s$ rest frame. 

Helicity conservation unequivocally determines the quantum state describing the spin of the two massive spin-1 particles, which is then pure for any values of the involved helicity amplitudes~\cite{Fabbrichesi:2023cev,Fabbrichesi:2023jep} and can be written as
\bea
|\Psi \rangle &=& \frac{1}{\sqrt{|\mathcal{M}|^2}} \Big[  w_{++}\, \ket{++} 
  +  w_{00} \, \ket{00} +  w_{--}\, \ket{--} \Big] \, ,
\label{pure}
\eea
with
\be
|\mathcal{M}|^2= |w_{00}|^2 + |w_{++}|^2 + |w_{--}|^2 \, .
\ee
The relative weight of the transverse components $ \ket{++}$ and $ \ket{--} $ with respect to the longitudinal one, $\ket{00}$, is controlled  by the conservation of angular momentum. The helicity density matrix $\rho = |\Psi \rangle \langle \Psi |$ for the pure state $|\Psi \rangle $  in \eq{pure}, is expressed in terms of the helicity amplitudes as\footnote{Or choice for the representation of the dimension-9 polarization basis of the 2-particle system is:
\begin{equation}
  \ket{+-} = 
  \begin{pmatrix}
   1 \\ 0 \\ 0\\ \vdots \\ 0 
  \end{pmatrix}\,,
  \quad
  \ket{+0} = 
  \begin{pmatrix}
   0 \\ 1 \\ 0\\ \vdots \\ 0 
  \end{pmatrix}\,,
  \quad
  \ket{++} = 
  \begin{pmatrix}
   0 \\ 0 \\ 1\\\vdots \\ 0 
  \end{pmatrix}\,,
  \quad
  \dots\,,
  \quad
  \ket{-+} = 
  \begin{pmatrix}
   0 \\ 0 \\ 0\\ \vdots \\ 1 
  \end{pmatrix}\,.
  \quad
\end{equation}
}
\be
\small
\rho=  \frac{1}{|\mathcal{M}|^2}\, \begin{pmatrix} 
  0 & 0 & 0 & 0 & 0 & 0 & 0 & 0 & 0  \\
  0 & 0 & 0 & 0 & 0 & 0 & 0 & 0 & 0  \\
  0 & 0 &  w_{++} \,w_{++}^* & 0 &  w_{++}  \,w_{00}^*& 0 &  w_{++}\, w_{--}^*& 0 & 0  \\
  0 & 0 & 0 & 0 & 0 & 0 & 0 & 0 & 0  \\
  0 & 0 & w_{00} \,w_{++}^* & 0 & w_{00} \, w_{00}^*  & 0 & w_{00} \, w_{--}^*& 0 & 0  \\
  0 & 0 & 0 & 0 & 0 & 0 & 0 & 0 & 0  \\
  0 & 0 &  w_{--} \,w _{++}^*& 0 &  w_{--}\, w_{00}^*& 0 &   w_{--}\,w_{--}^*& 0 & 0  \\
  0 & 0 & 0 & 0 & 0 & 0 & 0 & 0 & 0  \\
  0 & 0 & 0 & 0 & 0 & 0 & 0 & 0 & 0  \\
\end{pmatrix} \, .
\label{rhoBVV}
\ee

\subsection{Entanglement}

Quantifying the amount of entanglement in a quantum system is generally quite a challenging task. The complexity of the problem increases with the  dimensionality of the system and, often, partial answers are all that is available when dealing with systems more complicated than qubits. We refer the Reader to Refs.~\cite{Horodecki:2009zz,Barr:2024djo} for more details pertaining to entanglement monotones and measures.  

On general grounds, the state of a bipartite system is called entangled if it is \underline{not} \textit{separable}, that is if it \underline{cannot} be written in the form
\begin{equation}
  \rho_{\rm sep} = \sum_i p_i \, \rho_A^i \otimes \rho_B^i\,, \quad \text{with}\,\sum_i p_i = 1\,,
\end{equation} 
where $\rho_A^i$ and $\rho_B^i$ are the density operators describing, with probability $p_i$, the subsystems of the bipartite system. This implies that a pure state $\ket{\psi}$ is entangled if it is \underline{not} a product state, i.e. if it \textit{cannot} be written as $\ket{\psi} = \ket{\psi_1} \otimes \ket{\psi_2}$ with $\ket{\psi_1}$ and $\ket{\psi_2}$ being the pure states of the composing subsystems. 

If the bipartite system is in pure state, as that in \eq{rhoBVV}, it is possible to quantify its entanglement by computing the \textit{entropy of entanglement}
\be
{\cal E}= -\tr(\rho_A \log{\rho_A})=-\tr(\rho_B \log{\rho_B})\, ,
\label{entropy}
\ee
which is given by the von Neumann entropy~\cite{Horodecki:2009zz} of either of the two subsystems with reduced density operators $\rho_A$ and $\rho_B$. The latter are obtained from the density operator of the whole bipartite system by taking the \textit{partial trace} over the subsystem $B$ and $A$, respectively. This is formally defined so that
\begin{equation}
  \tr(\rho_B M) = \tr((\mathbb{1}_A\otimes M) \rho)\,.
\end{equation}

The von Neumann entropy of the reduced density operator is a true entanglement measure satisfying the property 
\be
0 \le {\cal E} \le \log{d}\, ,
\ee
where $d$ is the dimensionality of the subsystem --- $d = 3$ for a two-qutrit system.  The first equality holds if and only if the bipartite state is separable, the second inequality saturates if the bipartite state is maximally entangled. 

In the following we employ the entropy of entanglement to quantify the amount of entanglement present in the spin correlations of the $\phi$ meson pairs produced by the decaying $B^0_s$ mesons. 

\subsection{Bell inequality violation}
Local deterministic theories satisfy an inequality, known as the Bell inequality~\cite{Horodecki:1995340,Brunner:RevModPhys.86.419,Collins:2002sun,Kaszlikowski:PhysRevA.65.032118}, pertaining to correlated measurements performed by two independent observers on the subsystems $A$ and $B$ of the bipartite system. Crucially, the inequality may be violated by the statistical predictions of quantum mechanics due to the non-local character of entanglement and experimental tests of the inequality may then be used to discriminate between the two competing descriptions.

A re-formulation of the original Bell inequality adapted to the case of a bipartite system made of two qutrits\footnote{
  We refer the reader to~\cite{Horodecki:2009zz,Barr:2024djo} for a general discussion of the formulation of the Bell inequality for bipartite qubit systems and higher dimensional systems.} is the Collins,
Gisin, Linden, Massar, and Popescu (CGLMP) inequality~\cite{Collins:2002sun,Kaszlikowski:PhysRevA.65.032118, Horodecki:1995340,Brunner:RevModPhys.86.419}. In order to write the inequality explicitly, consider again the two qutrits A and B of the bipartite system. For the qutrit A, select two spin measurement settings, $\hat{A}_1$ and $\hat{A}_2$, which correspond to the projective measurement of two spin-1 observables having each three possible outcomes
$\{0, 1, 2\}$. Similarly, the measurement settings and corresponding observables for the qutrit B are $\hat{B}_1$ and $\hat{B}_2$. One can then construct the correlation measure, $\II$, given by the combination~\cite{Collins:2002sun}:
\bea
\II &=& P(A_1 = B_1 ) + P(B_1 = A_2 + 1) + P(A_2 = B_2) + P(B_2 = A_1)
\nonumber \\
&-&P(A_1 = B_1 - 1) - P(A_1 = B_2)- P(A_2 = B_2 - 1) - P(B_2 = A_1 - 1) \,,
\label{I3}
\eea
in which $P (A_i = B_j + k)$  denotes the probability that the outcome $A_i$ (for the measurement of $\hat{A}_i$) and $B_j$ (for the measurement of $\hat{B}_j$), with $i, j$ either 1 or 2, differ by $k$ modulo 3. The correlations computed within deterministic local models inevitably yield $\II \le 2$, while the bound can be violated by computing the above correlation measure with the rules of quantum mechanics.

Within the density matrix formalism, the above quantity can be written as an expectation value of a suitable Bell operator $\Bell$
\cite{Collins:2002sun}
\be
\II=\tr(\rho \Bell)\,, 
\ee
the explicit form of which will depend on which of the four operators $\hat{A}_i$, $\hat{B}_i$, with $i=1$ or $2$, are utilized in the test. Consequently,
for a set density matrix it is possible to enhance the violation of the Bell inequality through a specific choice of these operators. Regardless of the choice, the numerical value of the observable is bound to be less than or equal to $4$ if quantum mechanics is to hold true. Within the choice of measurements defining the Bell operator there is furthermore still freedom of modifying the measured observables through local unitary transformations. Correspondingly, the Bell operator undergoes the transformation:
\be
\Bell \to (U\otimes V)^{\dag} \cdot \Bell\cdot (U\otimes V)\ , \label{uni_rot}
\ee
where $U$ and $V$ are independent $3\times 3$ unitary matrices. This procedure effectively corresponds to local changes of the bases used in the two measurements and in the following we make use of it to maximize the value of ${\cal I}_3$ obtained with the Bell operator of Ref.~\cite{Acin:2002zz}.

\section{Results} 
We proceed now with the results pertaining to entanglement and Bell inequality violation in the process $B^0_s\to \phi \phi$. Corresponding results for other relevant $B^0_d$ and $B^0_s$ meson decays into two spin-1 mesons will also be mentioned.

In order to compute the helicity density matrix in \eq{rhoBVV} for the
$B^0_s\to \phi \phi$ decay process, we use the results of the analysis in Ref.~\cite{LHCb:2023exl} reported in table~\ref{tab:res}. These results provide the measurements of the two complex polarization amplitudes $A_{0}$ and $A_{\perp}$, including the relative phases. The polarization amplitudes are complex mostly because of the final-state interactions~\cite{Beneke:1999br} and the remaining amplitude, $|A_{\parallel}|$, can be derived by the condition 
$|A_{0}|^{2}+|A_{\perp}|^{2}+|A_{\parallel}|^{2}=1$. In our work we have set $\delta_0=0$ because there are only two physical phases entering in the polarizations amplitudes, namely $(\delta_{\perp}-\delta_0)$ and $(\delta_{\parallel}-\delta_0)$. The correlations among the amplitude and phase uncertainties, given in the supplementary material of Ref.~\cite{LHCb:2023exl}, are reported in table~\ref{tab:corr}.

\begin{table}[ht]
  \begin{center}
  \begin{tabular}{cc}
  \hline
  Parameter & Result\\
  \hline
  $|A_{0}|^{2}$ &   0.384 $\pm$ 0.007 $\pm$ 0.003 \\ 
  $|A_{\perp}|^{2}$ & 0.310 $\pm$ 0.006 $\pm$ 0.003 \\
  $\delta_{\parallel}$ [rad] & 2.463 $\pm$ 0.029 $\pm$ 0.009\\
  $\delta_{\perp}$ [rad] & 2.769 $\pm$ 0.105$\pm$ 0.011\\
\hline
\end{tabular}
 \end{center}
    \caption{Values of the complex polarization amplitudes and relative phases utilized in this work. The central values and the corresponding statistical and systematic errors are taken from Ref.~\cite{LHCb:2023exl}.}
  \label{tab:res}
\end{table}

\begin{table}[ht]
  \begin{center}
  \begin{tabular}{lcccc}
  \hline
& $|A_0|^2$  & $|A_{\perp}|^2$ &  $\delta_{\parallel}$ & $\delta_{\perp}$ \\ \hline
$|A_0|^2$   & 1 &  -0.342&   -0.007&   0.064 \\
$|A_{\perp}|^2$ &   &   1    &    0.140&   0.088 \\
$\delta_{\parallel}$ &  &  & 1    &0.179 \\
$\delta_{\perp}$  &  &  &   & 1 \\
\hline
\end{tabular}
 \end{center}
    \caption{ Correlation matrix for the measurements utilized in this work, originally presented in the supplementary material of Ref.~\cite{LHCb:2023exl}.}
  \label{tab:corr}
\end{table}

The helicity amplitudes are mapped into the polarization amplitudes typically used in experimental analysis~\cite{LHCb:2013vga,Belle:2005lvd,LHCb:2018hsm,ATLAS:2020lbz,LHCb:2023exl} through the relations
\be
\frac{w_{00}}{\abs{\mathcal{M}}}= A_{0} \,, \quad \frac{w_{++}}{\abs{\mathcal{M}}}  = \frac{ A_{\parallel}+A_{\perp}}{\sqrt{2}}\, , \quad
\frac{w_{--}}{\abs{\mathcal{M}}}  = \frac{ A_{\parallel} -A_{\perp}}{\sqrt{2}} \, ,
\ee
which we utilize to reconstruct the density matrix in \eq{rhoBVV}. We then can employ \eq{entropy} to quantify the amount of entanglement present in the spin correlations of the $\phi$ mesons produced in $B^0_s\to \phi \phi$ decays, finding~\cite{Fabbrichesi:2023idl}
\be
\boxed{
\mathscr{E} = 0.734 \pm  0.037 \label{sig_ent}
}\,.
\ee
This result demonstrates that the presence  of
quantum entanglement in $B^0_s\to \phi \phi$ is established with a significance exceeding the 5$\sigma$ threshold (nominally 19.8$\sigma$). The computation of the quoted error was performed by propagating the experimental uncertainties while taking into account the correlations in table~\ref{tab:corr}.

It is interesting to see what happens to the entanglement when the strong phases, sourced by final state interactions, are neglected. In this case the corresponding result of \eq{sig_ent} for the entropy of entanglement at vanishing phases, $(\mathscr{E}_0)$, is
\be
\mathscr{E}_0 = 0.666 \pm  0.0028 \label{sig0_ent}
\, ,
\ee
and shows a decrease in entanglement. We therefore conclude that the effect of strong rescattering of the final state particles increases the entanglement of the dimeson system. The result is interesting as it allows to isolate the effect of these strong processes on quantum entanglement.

We compute next the correlation measure ${\cal I}_3$ in \eq{I3} and maximize the Bell inequality by means of the procedure detailed in \eq{uni_rot}. The maximum value of ${\cal I}_3^{\rm max}$ obtained is given by~\cite{Fabbrichesi:2023idl}
\be
\label{resi3}
\boxed{
{\cal I}^{\rm max}_3  = 2.525 \pm 0.064
}\, ,
\ee
which implies that the CGLMP inequality ${\cal I}_3 < 2$ is also violated with a significance exceeding the $5\sigma$ level (nominally 8.2$\sigma$). The error quoted in \eq{resi3} was computed by maintaining the two matrices $U$ and $V$ set to the values indicated by the maximization procedure. These are reported below with an accuracy of $1\%$ from the obtained numerical values: 
\be
U=\left(
\begin{array}{ccc}
 -\frac{1}{25}-\frac{21 i}{37} & -\frac{8}{33}+\frac{59
   i}{89} & -\frac{11}{43}-\frac{i}{3} \\ \\
 \frac{104}{207}+\frac{13 i}{41} & 0 &
   -\frac{29}{36}+\frac{i}{60} \\ \\
 -\frac{9}{17}+\frac{5 i}{24} &
   -\frac{12}{17}-\frac{i}{30} & -\frac{7}{29}+\frac{11
   i}{32} \\
\end{array}
\right)\, , ~~
V=\left(
\begin{array}{ccc}
 -\frac{3}{35}+\frac{16 i}{39} & -\frac{5}{32}+\frac{20
   i}{29} & \frac{11}{21}-\frac{2 i}{9} \\ \\
 -\frac{7}{13}-\frac{43 i}{72} & 0 &
   \frac{3}{22}-\frac{15 i}{26} \\ \\
 \frac{5}{12}-\frac{i}{24} & -\frac{31}{44}+\frac{2
   i}{35} & -\frac{10}{27}-\frac{13 i}{30} \\ \\
\end{array}
\right)\,.
\ee
Notice that the value of ${\cal I}_3$ obtained without optimization is $
{\cal I}_3  = 0.86 \pm 0.043$, which implies no violation of CGLMP inequality. As we can see, the optimization procedure is here crucial to unveil the violation of the Bell inequality.

We conclude the section by listing,  in table~\ref{other:results}, the results obtained for the entropy of entanglement and the optimized Bell operator using the data of other $B$-meson decays.

\begin{table}[h]
  \centering
\begin{tabular}{l  c c}\\
  Process& $\boldsymbol{\mathscr{E}}$ &${\cal I}_{3}^{\rm max}$ \\[+0.2cm]
\hline\\
$ \; \;\boldsymbol{B^{0}_d\to J/\psi  \,\K}$~\cite{LHCb:2013vga}\hspace{1cm} & $\qquad 0.756 \pm 0.009 \qquad$   & $2.548 \pm 0.015$ \\[+0.2cm]
$\;\;\boldsymbol{B^0_d\to \phi \,\K}$~\cite{Belle:2005lvd} & $0.707\pm 0.133^{\ast}$    &  $2.417\pm0.368^{\ast}$\\[+0.2cm]
$\;\;\boldsymbol{B^{0}_d\to \rho \, \K }$~\cite{LHCb:2018hsm} &  $0.450\pm 0.077^{\ast}$  &  $2.208 \pm 0.151^{\ast}$\\[+0.2cm]
$\;\;\boldsymbol{B_s^{0}\to J/\psi  \,\phi}$~\cite{ATLAS:2020lbz} & $0.731\pm0.032$    &  $2.462\pm 0.080 $\\
\\
\end{tabular}
\caption{ Amount of entanglement and optimized expectation value of the Bell operator for other $B$-meson decays~\cite{Fabbrichesi:2023idl}. An asterisk indicates that the correlations in the uncertainties of the helicity amplitudes are not given in the corresponding reference and therefore only an upper bound on the  propagated uncertainty can  be  computed.}
\label{other:results}
\end{table}

\section{Discussion}
Although the $B^0_d\to \JP~\K$ decays result in a larger evidence for the detection of the Bell inequality violation -- see table~\ref{other:results}, the fact that the $\JP$ has a much longer lifetime than the $\K$ allows for the causal propagation of signals among the two subsystems prior to the conclusion of the measurement act. If this were to happen, it would invalidate one of the hypothesis of the Bell test, as well as its conclusions for the related locality loophole cannot be disregarded. 

To understand the issue, notice that particles often act as their own polarimeters, in that the directions of the decay products reveal the orientation of the progenitor particle spin vector, allowing for its reconstruction. The locality loophole exploits here the fact that once the $\K$ decays ``measuring", so to say, its own spin, a causal interaction could inform the $\JP$ particle about the specifics of this measurement, thereby conditioning the impending decay and second measurement. Within an idealized Bell test, the situation is analogous to that in which one of the two observers involved, say Alice, informs the other, Bob, of the selected measurement settings and of the outcomes they obtained. This would clearly invalidate the test as Bob could simply fake the presence of correlation by calibrating their own measurements to the results that Alice obtained. How can we then trust that Bob's results are fair? One simple solution is to place Alice and Bob outside of each other's light cone, preventing every possible form of causal communication. As we argue below, this is exactly what happens in case of the $B$ meson decaying into two $\phi$ mesons.  

Because the final states comprise identical particles, it is plausible that configurations where the decays of the $\phi$ mesons take place basically simultaneously could be more easily achievable. Still, decays are inherently stochastic processes that take place with an exponential spread and so one must verify that a large majority of these decays do take place at space-like separations. If so, we could safely argue that the locality loophole cannot significantly affect the conclusions of our Bell test.

Consider then two particles, A and B, produced together in a decay process. The particle B may decay within the future cone of the particle A either because of a longer lifetime or because of the random spread in its decay time. Regardless, the relative velocity $v$ with which the pair flies apart will be sufficiently large to create a space-like separation between the forthcoming decays if 
\be
\frac{|t_1-t_2|c}{(t_1+t_2)v}<1\, ,
\label{locality}
\ee
in which $t_1$ and $t_2$ are the instants at which the corresponding particles decay. This separation condition prevents a possible communication through local interactions, such as usual the electromagnetic interaction mediated by photons, and therefore guarantees that the locality loophole is closed~\cite{bell2004speakable}. Experimentally, \eq{locality} can be enforced by imposing a cut on the momenta of the produced particles apt to reject the pairs which do not separate fast enough. If the amount of available data after the cut remains large enough to guarantee that a fair sample of events can be collected (thereby addressing the detection loophole), this procedure would not affect the significance of the Bell test under consideration.

In the case of the $B^0_s\to \phi \phi$ decay we can estimate the fraction of events separated by a space-like interval by performing a simple Monte Carlo simulation in which the decay times come from the usual exponential distribution regulated by the $\phi$ decay width. To this purpose we model the lifetime of each $\phi$ particle in a function $t_i=F(x_i)$, $i=1$ or 2, depending on the random variable $x_i$ that we restrict to the range $0\le x_i\le 1$.  The explicit form of the $F(x_i)$ function is
\bea
F(x_i)&=&-\frac{\tau_{\phi}\log\left(x_i\right)}{\sqrt{1-\beta^2}}\, ,
\eea
with $\tau_{\phi}$ being the $\phi$ meson life-time, $\beta=\sqrt{1-4 M_{\phi}^2/M_{B_s}^2}$ the $\phi$ velocity in the $B$ meson rest frame (in units of c), and $M_{\phi}$ and $M_{B_s}$ the $\phi$ and $B^0_s$ meson mass, respectively.  

As anticipated in the Introduction, by performing $10^5$ pseudo-experiments (corresponding to $10^5$ extractions of random values for the pair ($x_1$, $x_2$)),  we find that almost 95\% of the events are space-like separated and every form of local communication between the two parties is then automatically prevented. The procedure can be easily generalized to cases containing two different particles in final state.

Concerning the effects of strong interactions on the $\phi\phi$ bipartite system, we also see that the decays of the two $\phi$ mesons take place well outside of the range of the interactions ongoing at the time of their production, as well as of that of final-state interactions induced by virtual mesons exchanges. The former is a pure QCD effect due to gluon exchanges and is active on distances of about $\sim 3 \times 10^{-5}$ fm~\cite{Yamamoto:2008ze}.
The range of the latter, instead, is at most equal to $\lambda_{\pi} \simeq $ 1.5 fm if we consider an interaction mediated by one-pion exchange. When compared to the spatial separation between the two $\phi$ mesons decay vertices, which is typically of $\mathcal{O}(100) $ fm, these numbers indicate that no strong interaction exchange can happen between the two $\phi$ mesons at the time of their decays.

\section{Summary}

We review the possibility of detecting entanglement and the violation of Bell inequalities that it might entail by using the decays of $B$ mesons into pairs of vector mesons, specifically focusing on the $B^0_s$ decays that yield two $\phi$ mesons. Using data from the LHCb experiment, we show that the reconstructed spin correlations establish the presence of quantum entanglement in the bipartite system formed by the $\phi$ mesons. The analysis also ascertains the violation of the Bell inequality in these decays, thereby supporting the quantum mechanical nature of the strong and electroweak interactions involved. Additionally, the paper addresses potential experimental loopholes that could invalidate the results. In particular, we find that the locality loophole is closed for the considered $\phi \phi$ final state as the two particles decay separated by a space-like interval.

\section*{Acknowledgements}
We thank M. Fabbrichesi for useful discussions.
E.G. acknowledges the Department of Theoretical Physics of CERN for the kind hospitality during the preparation of this work. L.M. acknowledges the Estonian Research Council for supporting the work with the  grants PRG803, RVTT3 and with the CoE program grant TK202 ``Fundamental Universe'’.

\bibliographystyle{JHEP}

\bibliography{review-Bmeson.bib}

\end{document}